\documentstyle[12pt,epsfig]{article}

\newlength{\dinwidth}
\newlength{\dinmargin}
\def\lapproxeq{\lower .7ex\hbox{$\;\stackrel{\textstyle<}{\sim}\;$}}
\def\gapproxeq{\lower .7ex\hbox{$\;\stackrel{\textstyle>}{\sim}\;$}}
\def\xg{x_{\gamma}}
\def\alps{\alpha_s}
\def\mxt2{\tilde{M}_X^2}
\def\myt2{\tilde{M}_Y^2}
\newcommand{\xpom}{x_{_{\rm I\!P}}}

\begin{document}
\titlepage
\begin{flushright}
MC-TH-98/7 \\
MAN/HEP/98/2 \\
April 1998
\end{flushright}

\begin{center}
\vspace*{2cm}
{\Large \bf 
Double Diffraction Dissociation at High $t$
} \\
\vspace*{1cm}
B.E.~Cox and J.R.~Forshaw \\
\vspace*{0.5cm}
Department of Physics and Astronomy, \\
University of Manchester,\\
Manchester, M13 9PL, England.
\end{center}

\vspace*{5cm}
\begin{abstract}
Diffractive scattering in the presence of a large momentum transfer is an
ideal place to study the short distance rapidity gap producing mechanism.
Previous studies (experimental and theoretical) in this area have focussed on 
gaps between jets and on high-$t$ vector meson production. We propose the
measurement of double dissociation at high-$t$. We examine the numerous 
advantages to studying this more inclusive process and conclude that it is 
an ideal place to study short distance diffraction.  
\end{abstract}

\newpage

\section{Introduction}
In order to develop our understanding of diffractive processes, and in
particular to understand them within the framework of QCD, it has been
recognised that it is useful to focus in on a special class of diffractive
phenomena: those that are driven by short distance physics. Large momentum 
transfer processes are especially good
filters of short distance diffraction \cite{FS}. When we speak of large 
momentum transfer, we mean that the square of the four-momentum transferred 
across the associated rapidity gap, $-t$, is large. So, with this definition, inclusive 
small-$x$ deep inelastic scattering is not a large momentum transfer process 
(it is at $t=0$) whereas the `rapidity gaps between jets' process suggested for
study by Mueller \& Tang \cite{MT} and measured at HERA \cite{gapHERA} and 
the TEVATRON \cite{gapTEV} colliders is. Quasi-elastic vector meson 
production at high $p_t$ is another example of a high-$t$ process which has
been studied theoretically \cite{meson} and by experiment \cite{mesonexp}.

Let us begin by reviewing the current situation regarding the `gaps between
jets'. Recall that one looks at the dependence of the production rate 
for events which contain a pair of jets and which have a rapidity gap between
the jets, as a function of the gap size. The transverse momentum of the jets
then makes sure that the gap producing mechanism is driven by small,
i.e. $\lapproxeq 1/p_t$, distances. Consequently, one can apply perturbative 
QCD and aim to really test the theory. On the theoretical side, the
calculations have been performed within the leading logarithmic approximation
(LLA) of BFKL \cite{BFKL} and predict an exponentially strong rise of the 
cross-section with increasing gap size. However, we note the LLA is not 
sufficient for a precise quantitative test of the theory \cite{NLO}. 
In order to access the most interesting physics, it
is necessary that data be collected at large rapidity gaps ($\gapproxeq 4$).
Unfortunately, the demand to observe a pair of high-$p_t$ jets within the
detectors severely restricts the experimental reach. The TEVATRON experiments
have collected data with gaps of up to 6 units between the jet centres, i.e. 4 units 
between the jet edges, whilst at HERA the experiments can only reach gaps of 
about 4 units between the jet centres. At HERA, this problem 
is somewhat circumvented by looking at high-$t$ vector meson production 
\cite{meson}, but the rate is a low (although in time sufficient data
will be collected) and one is limited to investigating the production of 
systems with fixed invariant mass. One also has the further complication 
that one needs to understand how the meson gets produced.

It is possible to avoid many of the above drawbacks by studying the more
inclusive process where the incoming beam particles each dissociate, 
producing systems $X$ and $Y$ which are far apart in rapidity, in the
presence of a large momentum transfer. The two processes discussed in the
previous paragraph form subsets of this more inclusive measurement. 
Our subsequent studies will focus mainly on double dissociation at HERA, i.e.
in photon-proton interactions. We expect similar results for double
dissociation at the TEVATRON. 

\begin{figure} 
\centerline{\epsfig{file=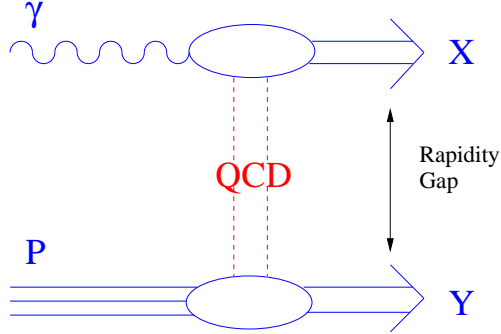,height=6.5cm,angle=270}}
\caption{The double diffraction dissociation process.}
\label{diffplot}
\end{figure}

Figure \ref{diffplot}  shows the double dissociation process of our study. 
The photon
dissociates to produce the system $X$, which travels in the backward 
direction whilst the proton dissociates to produce the system $Y$, which
heads off in the forward direction. By not requiring to see a pair of jets in 
the detector we immediately open up the reach in rapidity by, 
as we shall shortly see, a very significant amount. We shall focus on photoproduction
with a tagged electron (i.e. $Q^2 \lapproxeq 10^{-4}$ GeV$^2$).\footnote{
It is also very interesting to make the measurements at high $Q^2$.} 
An additional advantage is that the asymmetry of the beam energies at HERA `draws' the 
system $X$ out of the backward beam hole making a measurement of its four-vector
(and hence its invariant mass, $M_X$) possible.
The entire kinematics of the event are determined by the measurement of the 
scattered electron (giving the four-vector of the photon) and the four-vector
of the system $X$. 

In order to select the events we do not propose that a gap cut be
made. We follow the H1 procedure, wherein one uses the whole of the
inclusive tagged photoproduction sample and, for each event, searches for
the largest gap in the event \cite{H1diff}. This gap is then used to define the systems
$X$ and $Y$, i.e. all particles backward of the gap go into system $X$ and
all those forward of the gap go into system $Y$. Events with gaps are
characterised by $M_X, M_Y \ll W$ and so we choose to reduce the sample 
by insisting that $M_Y \le M_{Y{\rm cut}}$. However we note that,
given the momentum transfer four-vector, one can deduce $M_Y$ and hence 
that it is possible to measure the fully differential cross-section 
$d^4 \sigma / (dW dt \ dM_X dM_Y)$. 
We stress that we are not defining the sample in terms of a rapidity gap cut.
This way of defining the sample has the very important advantage that
it is easy to implement theoretically and is quite insensitive to
hadronisation effects which define the edges of the systems $X$ and $Y$.

The remainder of the paper is organised as follows. We begin by presenting the
relevant cross-section formulae. We then turn to Monte Carlo in order 
to examine issues relating to non-colour-singlet
exchange and possible non-perturbative effects. We close by drawing our
conclusions and alerting the reader to necessary future studies.  
 
\section{Kinematics and the double dissociation cross-section} 
Let $q_{\mu}$ be the four-momentum transfer. It
has the following Sudakov decomposition:
$$
q_{\mu}= \alpha P_{\mu} + \beta Q_{\mu} + q_{t,\mu}
$$
where $P$ and $Q$ label the four-momenta of the incoming proton and photon. (We
assume $P^2 = Q^2 = 0$.) In the Regge limit, we have
 $t \equiv q^2 \approx q_t^2$ and can write
\begin{eqnarray}
M_X^2 &\approx& t - \beta W^2 \nonumber \\
M_Y^2 &\approx& t + \alpha W^2
\end{eqnarray}
where $W^2 = (P+Q)^2 = 2 P \cdot Q$.

Now, in the limit $-t \gg \Lambda_{{\rm QCD}}^2$ we can factorise the long distance
physics into parton distribution functions and assume that the gap is produced by
a single elastic parton-parton scattering, i.e. we can write
\begin{equation}
\frac{d \sigma(\gamma p \to XY)}{d\xg dx_h dt} = F_{\gamma}(\xg,-t) F_p(x_h,-t)
\frac{d \hat{\sigma}(qq \to qq)}{dt}
\end{equation}
where $F_i(x,\mu^2) = \frac{9}{4} g_i(x,\mu^2) + \Sigma_i(x,\mu^2)$ is a colour weighted 
sum over gluon and quark parton density functions for beam particle $i$. 
The fraction of the
beam momentum carried by the struck parton is $x$ and $\mu$ is the factorisation
scale. The perturbative QCD dynamics lies in the cross-section for elastically
scattering a pair of quarks: $d \hat{\sigma}(qq \to qq)/dt$. In the leading
logarithm approximation, this is given by exchanging a pair of interacting
reggeised gluons \cite{BFKL}. In the asymptotic limit $\xg x_h W^2 \gg -t$ the
cross-section can be computed analytically \cite{MT}:
\begin{equation}
\frac{d \hat{\sigma}(qq \to qq)}{dt} = (\alps C_F)^4 \frac{2\pi^3}{t^2}
\frac{e^{2 \omega_0 Y}}{(7 \alps N_c \zeta(3) Y)^3}
\end{equation}
where $$Y = \ln \left(\frac{\xg x_h W^2}{-t} \right)$$ and $$\omega_0 = N_c \ 
4 \ln 2 \ \frac{\alps}{\pi}.$$
This result has already been coded into HERWIG \cite{HERWIG}, 
as has the exact leading logarithm
expression \cite{MH}. 

An alternative model for the gap producing mechanism is to assume that
two massive gluons are exchanged, in which case
\begin{equation}
\frac{d \hat{\sigma}(qq \to qq)}{dt} = \frac{\pi}{16 t^2} \alps^4  \frac{1}{\Delta^2}
\ln^2 \frac{\Delta+1}{\Delta-1},
\end{equation}
where
\begin{equation}
\Delta = (1 - 4 M^2/t)^{1/2}
\end{equation}
and $M$ is the gluon mass. We have also incorporated this process into HERWIG.

The incoming partons are collinear with the beam particles whilst, after the
scatter, they carry transverse momentum $\sqrt{(-t)}$. These partons are the seeds
of the jets which are produced in the gaps between jets measurement. Putting these
outgoing partons on shell gives
\begin{eqnarray}
\xg = \frac{-t}{-\beta W^2} &=& \frac{-t}{\mxt2} \nonumber \\ 
x_h = \frac{-t}{\alpha W^2} &=& \frac{-t}{\myt2}.
\end{eqnarray}
We have defined $\mxt2 \equiv M_X^2 - t$ and $\myt2 = M_Y^2 -t$.
These kinematic relations between the parton momentum fractions and the physical
observables allow us to write
$$ \mxt2 \myt2 \frac{d \sigma}{\mxt2 \myt2} = \xg x_h \frac{d \sigma}{d\xg dx_h}.$$

For the purposes of this paper, we will assume that $M_Y$ is confined to be
smaller than some value. Recall that the Regge limit demands that $W \gg M_X, M_Y$ and
hence that there be a rapidity gap between the systems $X$ and $Y$.\footnote{
The rapidity gap between systems $X$ and $Y$ is equal to $\ln(\xg x_h W^2/(-2t))$.}
In particular we take $M_Y < 10$ GeV. Although it would be very interesting to
look at variations with $M_Y$. 

We shall also use the popular notation
$$ \xpom \equiv - \beta = \frac{\mxt2}{W^2}.$$ In which case
\begin{equation}
\frac{d \sigma(\gamma p \to XY)}{d\xpom dt} = \frac{1}{t^2} \int dM_Y^2 \
(\xg x_h W)^2 F_{\gamma}(\xg,-t) F_p(x_h,-t) \frac{d \hat{\sigma}(qq \to qq)}{dt}.
\end{equation}

Sitting at fixed $M_X^2$ and $t$ the $\xpom$ dependence is determined completely by the
hard partonic scattering. For example, if 
\begin{equation}
\frac{d \hat{\sigma}(qq \to qq)}{dt} \sim e^{2 \omega_0 Y}   \label{test}
\end{equation}
then
\begin{equation}
\frac{d \sigma(\gamma p \to XY)}{d\xpom dt} \sim \left( \frac{1}{\xpom} 
\right)^{2 \omega_0+1}.
\end{equation}
Alternatively, one can look at the $\xpom$ dependence at fixed $W$ and $t$. In which case
the $\xpom$ dependence is dependent upon the photon parton densities, i.e.
if the hard scatter is given by (\ref{test}) then it follows that
\begin{equation}
\frac{d \sigma(\gamma p \to XY)}{d\xpom dt} \sim \left( \frac{1}{\xpom} 
\right)^{2 \omega_0+2} F_{\gamma}(\xg,-t) \label{fixWeq}
\end{equation}
where $\xg = -t/(\xpom W^2)$.

\begin{figure}[t] 
\centerline{\epsfig{file=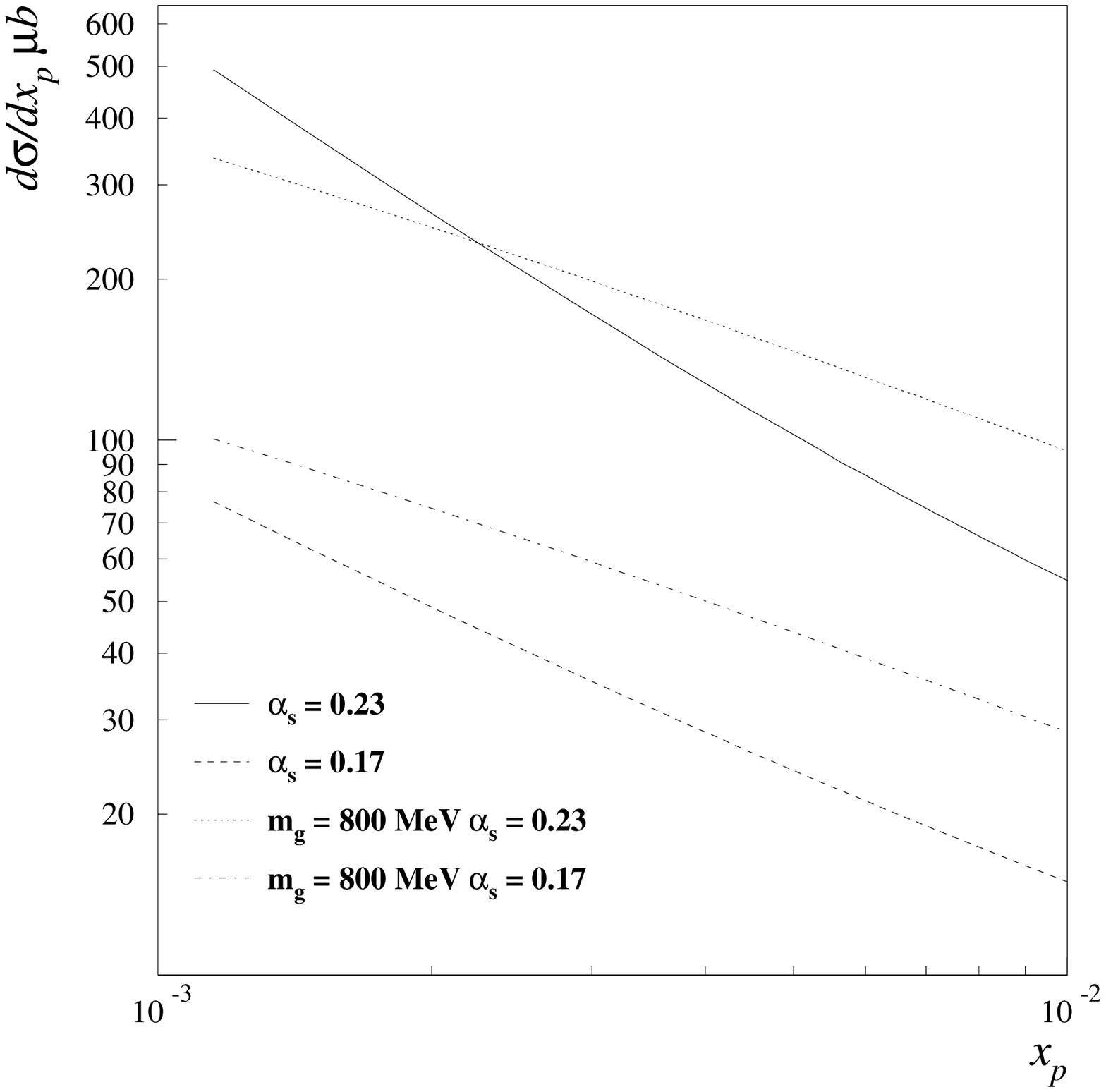,height=8cm,%
bbllx=28pt,bblly=143pt,bburx=542pt,bbury=650pt}}
\caption{Comparison of the BFKL (solid and dashed lines) and massive gluon (dotted and
dashed-dotted lines) predictions.}
\label{models}
\end{figure}

In Fig.\ref{models} we compare the massive gluon and BFKL calculations of the
$\gamma p$ cross-section (all the cross-sections we show will be for $\gamma p$
interactions). For the massive gluon calculation, we chose a gluon mass of 800 MeV. 
Decreasing the mass causes the cross-section to rise (it diverges at zero mass) but 
does not affect the shape of the $\xpom$ distribution; it remains significantly flatter
than the BFKL calculation. The cross-sections are shown for two different values of
the parameter $\alpha_s$. Increasing $\alpha_s$ steepens the BFKL distribution (since
$\omega_0 \sim \alpha_s$) but only affects the normalisation of the massive gluon 
distribution. 

For all our plots we take $W=250$ GeV, $-t>4$ GeV$^2$ and $M_Y < 10$ GeV. 
Unless otherwise stated, all our results
are derived using the BFKL calculation with $\alpha_s = 0.23$ and using the GRV-G HO
parton distribution functions of the photon \cite{GRVG} and the GRV 94 HO DIS proton 
parton density functions \cite{GRV94} as implemented in PDFLIB \cite{PDFLIB}. 
We choose to show plots at fixed $W$ (rather than fixed $M_X$ say) since this is 
the more natural scenario for a first measurement. 

\begin{figure}[b] 
\centerline{\epsfig{file=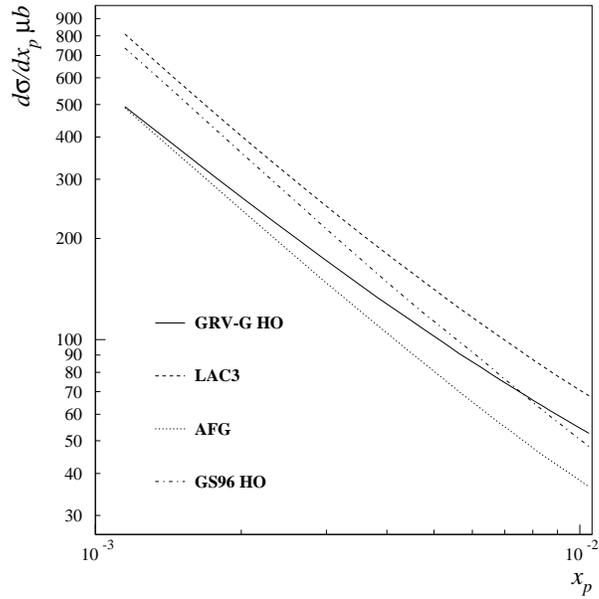,height=8cm,%
bbllx=28pt,bblly=143pt,bburx=542pt,bbury=650pt}}
\caption{Effect of varying the photon parton density functions.}
\label{pdf}
\end{figure}

In Fig.\ref{pdf} we show the effect of varying the photon parton density functions
\cite{GRVG,LAC,AFG,GS}. We note that the primary effect of varying the photon 
parton density functions is to shift the normalisation. In fact, for the BFKL
calculation, the $\xpom$ behaviour is quite close to $\sim (1/\xpom)^{2 \omega_0}$
independent of the photon parton density function (compare this with the 
expectation of (\ref{fixWeq})). Note that changing the proton parton density
functions does not affect the shape of the $\xpom$ distribution, since $x_h$
depends only upon $t$ and $M_Y$.

\section{Monte Carlo simulation}
Since the BFKL and massive gluon calculations have been implemented into HERWIG,
we are able to perform a simulation to hadron level which includes the background
expected from fluctuations in non-colour-singlet exchange photoproduction processes.
HERWIG is used to generate 2.5 pb$^{-1}$ of $ep$ data.
For each event with a tagged electron, the largest rapidity gap in the event is sought and
this is used to define the systems $X$ and $Y$. The kinematic cuts described
in the previous section are then applied.

\begin{figure}[t] 
\centerline{\epsfig{file=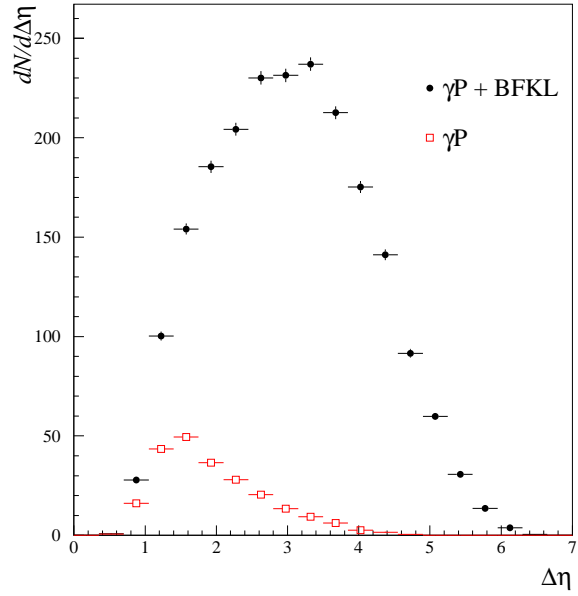,height=8cm,%
bbllx=0pt,bblly=145pt,bburx=538pt,bbury=650pt}}
\caption{Distribution of events in rapidity.}
\label{rap}
\end{figure}

In Fig.\ref{rap}  we show the distribution of our events as a function of 
the rapidity gap (only events which have $10^{-3} < \xpom < 10^{-2}$ are included). 
Those events which originate from non-colour-singlet exchange processes 
are also shown and can be seen to constitute a small fraction of the total. 
This effective depletion of the non-colour-singlet exchange contribution arises 
directly as a consequence of our insisting that $M_Y < 10$ GeV and 
$10^{-3} < \xpom < 10^{-2}$, i.e. that $M_X, M_Y \ll W$. In particular, no rapidity
gap cut is needed to select rapidity gap events.  

Since we are not requiring to see system $Y$ in the detector, it is now possible to 
select events with very large gaps. Also, since we are not requiring to see jets, many 
more events enter the sample than entered the gaps between jets measurement 
(this arises primarily because most events lie at the lowest values of $-t$). 
We wish to emphasise the magnitude of the gain in both statistics and reach in rapidity.
In particular, we note that the HERA gaps between jets data populate the 
$\Delta \eta < 2$ region and that, even in this restricted region, they constitute
only a small fraction of the total number of events.  

\begin{figure}[b] 
\centerline{\epsfig{file=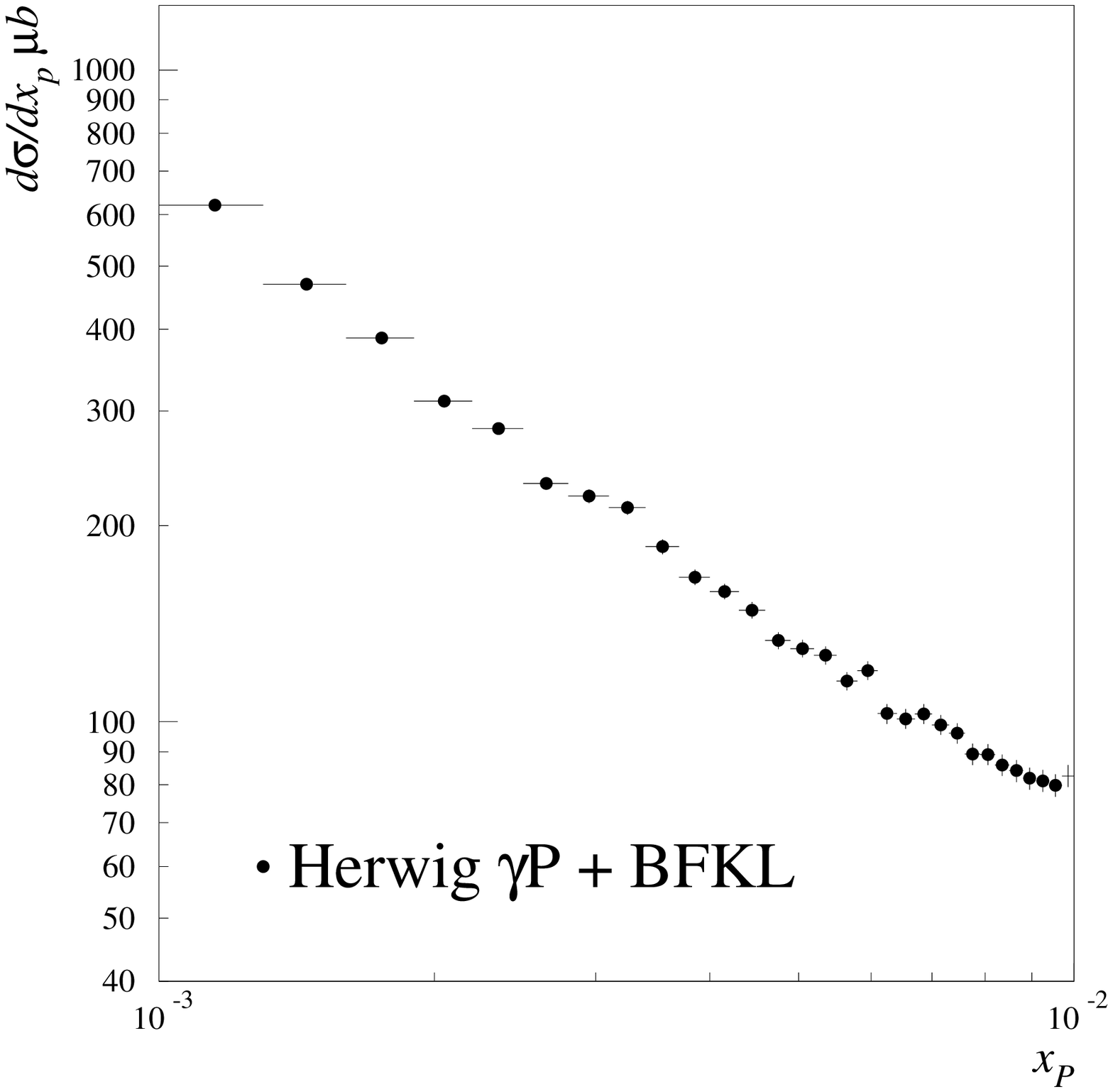,height=8cm,%
bbllx=0pt,bblly=145pt,bburx=538pt,bbury=650pt}}
\caption{The $\xpom$ distribution at fixed $W$.}
\label{xpom_fixW}
\end{figure}

\begin{figure}[t] 
\centerline{\epsfig{file=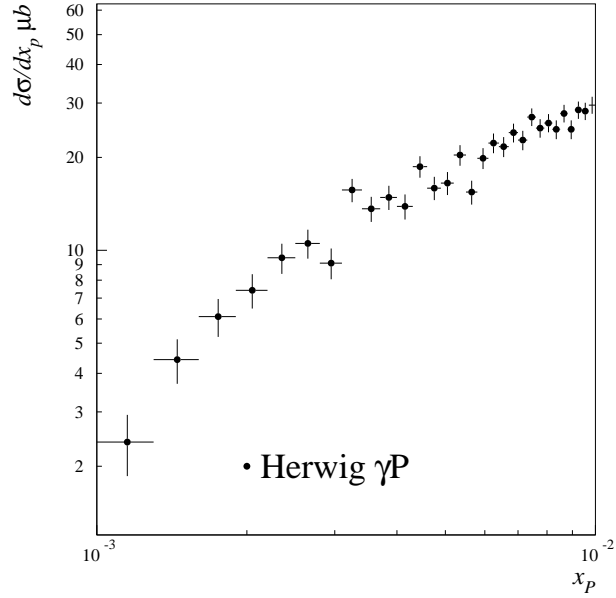,height=8cm,%
bbllx=0pt,bblly=145pt,bburx=538pt,bbury=650pt}}
\caption{The $\xpom$ distribution of the background due to non-colour-singlet exchange.}
\label{bgd}
\end{figure}

In Fig.\ref{xpom_fixW} we show the $\xpom$ distribution at fixed $W$. The errors
are typical of the statistical errors to be expected from analysis of data already
collected at HERA. The small contribution from non-colour-singlet exchange processes is
included. This is shown separately in Fig.\ref{bgd} and, as expected, it falls away 
exponentially as $\xpom$ falls. 

We expect that the data will be suitable for making very precise statements regarding the 
dependence of the gap production mechanism upon the gap size.

\begin{figure}[b] 
\centerline{\epsfig{file=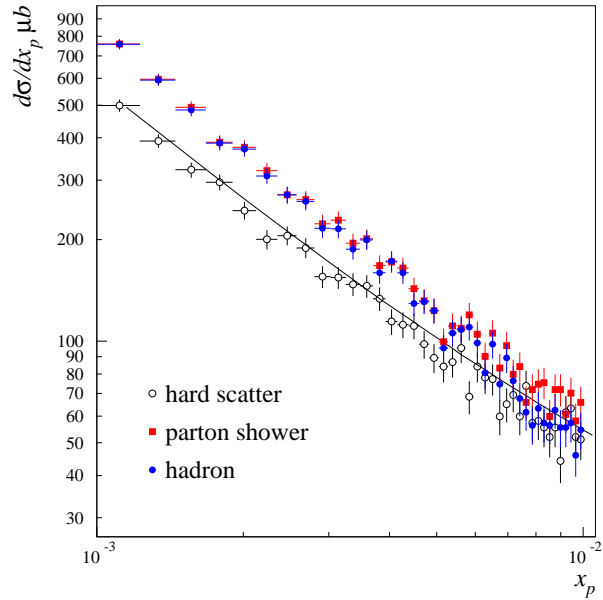,height=8cm,%
bbllx=0pt,bblly=145pt,bburx=538pt,bbury=650pt}}
\caption{Comparison of theory (solid line) with the Monte Carlo output from HERWIG.}
\label{link}
\end{figure}

In Fig.\ref{link} we compare the theoretical calculation with the output from
HERWIG. The open circles correspond to the cross-section obtained when $t$ is
constructed at the level of the hard scatter, i.e. it is the $t$ of the hard
BFKL subprocess. Of course this value of $t$ should be unique and equal to the
value extracted by summing the four-momenta of all outgoing hadrons in system
$X$ and subtracting from the four-momenta of the incoming photon. The solid
circles in Fig.\ref{link} show the effect of reconstructing $t$ at the hadron
level and clearly there is a significant shift relative to the hard scatter
points (which agree well with the theoretical calculation). The squares denote
the cross-section obtained by computing $t$ after parton showering but before
hadronisation. 

\begin{figure}[t] 
\centerline{\epsfig{file=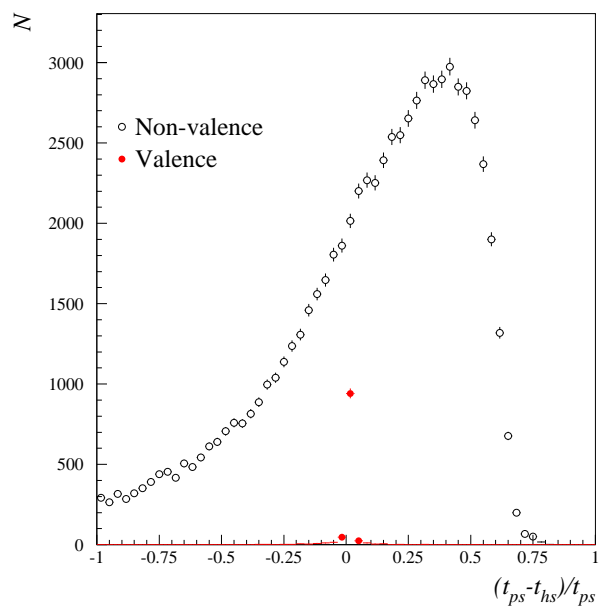,height=8cm,%
bbllx=0pt,bblly=145pt,bburx=538pt,bbury=650pt}}
\caption{Deviation of the reconstructed $t$ from the true value.}
\label{tplot}
\end{figure}

The fact that the reconstructed $t$ values are not all the same is clearly
demonstrated in Fig.\ref{tplot}. The number of events is plotted as a function
of the difference in $t$ between the value extracted after parton showering,
$t_{ps}$, and the value obtained from the hard scatter, $t_{hs}$. Note that 
there is no shift in $t$ for those events we have labelled `valence'. These
are events in which the parton entering the hard scatter from the proton is
either an up or a down quark and from the photon is not a gluon. All other
events are in the `non-valence' sample and it is here we see the systematic
shift to larger values of $-t$, i.e. $-t_{ps} > -t_{hs}$.  
This shift is a consequence of the way HERWIG develops the final state from
the seed partons in the hard scatter. In particular, since HERWIG generates 
the complete final state it necessarily has some model for the photon and proton remnants. 
For the proton, it will always evolve backwards from the hard scatter to a valence quark,
whereas for the photon evolution terminates with a quark--anti-quark pair.
Hence, in the `non-valence' sample at least one parton must be emitted off an
incoming parton. This is the source of the shift in $t$.

A shift to large $-t$ has the immediate consequence of systematically shifting the
$\xpom$ distribution to the right, as is clearly seen in Fig.\ref{link}. This is
easy to see since $\xpom = (M_X^2-t)/W^2$, so events which are generated with some
$t_{hs}$ are reconstructed, after parton showering, to have a larger $-t$ and hence
larger $\xpom$. In essence therefore, the difference between the theory and hadron level
results can be attributed to non-perturbative effects. The primary
feature of the effect is to systematically shift the $\xpom$ distribution without
significantly affecting its shape.

\section{Outlook}
We have proposed a new way of studying short distance diffraction. We believe this
measurement can be made by the HERA experimentalists. Things may not be so simple
at the TEVATRON, where one does not have the asymmetric beam energies to boost one
of the diffracted systems into the detectors. Our proposal optimises the statistics,
allows a wide reach in rapidity, has very small hadronisation corrections and very
little background. It is also easy to extract a cross-section for comparison with
theory. 

We have not discussed the issue of gap survival. This is a topic for further
discussion. For now let us point out that, should the gap survival probability 
decrease with increasing $W$ but be approximately constant at fixed $W$ (this
possibility is anticipated since the probability for secondary partonic interactions, 
which would destroy the gap, increases as the number density of partons increases and 
since the parton density is largest at small $x$, i.e. high $W$ \cite{MI}) then
the $\xpom$ at fixed $M_X$ and $\xpom$ at fixed $W$ distributions would provide
complementary information.  
 
\section*{Acknowledgements}
Thanks to Paul Newman and Mike Seymour for constructive discussions.


\end{document}